\documentclass[]{emulateapj}
\usepackage{graphicx}

\usepackage{color}
\definecolor{red}{rgb}{0.7,0.1,0.1}
\definecolor{blue}{rgb}{0.2,0.2,0.8}
\definecolor{green}{rgb}{0.1,0.6,0.1}

\def\cf{{\it cf.}}
\def\eg{{\it e.g.}}
\def\etal{{\it et al.}}

\def\ie{{\it i.e.}}

\def\pmb#1{\setbox0=\hbox{$#1$}%
  \kern-0.25em\copy0\kern-\wd0
  \kern.05em\copy0\kern-\wd0
  \kern-0.025em\raise.0433em\box0}

\long\def\Ignore#1{\relax}

\slugcomment{Revised version submitted to Apj Aug 16, 2016}

\shorttitle{Bar formation in growing disks}
\shortauthors{Berrier and Sellwood}

\begin{document}

\title{Mass Distribution and Bar Formation in Growing Disk Galaxy Models}

\author{Joel C. Berrier\altaffilmark{1}}
\altaffiltext{1}{Currently at Department of Physics,
University of Nebraska Kearney, 2401 11th Avenue, Kearney NE 68849}

\author{J. A. Sellwood}

\affil{Department of Physics and Astronomy, Rutgers University, 136
  Frelinghuysen Road, Piscataway, NJ 08854, USA}

\begin{abstract}
We report idealized simulations that mimic the growth of galaxy disks
embedded in responsive halos and bulges.  The disks manifested an
almost overwhelming tendency to form strong bars that we found very
difficult to prevent.  We found that fresh bars formed in growing
disks after we had destroyed the original, indicating that bar
formation also afflicts continued galaxy evolution, and not just the
early stages of disk formation.  This behavior raises still more
insistently the previously unsolved question of how some galaxies
avoid bars.  Since our simulations included only collisionless star
and halo particles, our findings may apply to gas-poor galaxies only;
however the conundrum persists for the substantial unbarred fraction
of those galaxies.  Our original objective was to study how internal
dynamics rearranged the distribution of mass in the disk as a
generalization of our earlier study with rigid spherical components.
With difficulty, we were able to construct some models that were not
strongly influenced by bars, and found that halo compression, and
angular momentum exchange with the disk did not alter our earlier
conclusion that spiral activity is largely responsible for creating
smooth density profiles and rotation curves.
\end{abstract}


%#####################################################################

%-----------------------------------------------------------
\section{Introduction} \label{sec:intro}
%-----------------------------------------------------------

Understanding the formation of galaxies is one of the grand challenges
of current astronomy \citep[\eg][]{SD15}.  In parallel with intense
observational programs, such as CANDELS \citep{Grog11, Koek11}, many
groups are pursuing simulations of cosmic structure formation that
attempt to follow the evolution of both dark matter and baryons from
{\it ab initio\/} conditions \citep[\eg][]{Gued11, Muns13, Snyd15,
  Scha15}, while others employ semi-analytic techniques
\citep[\eg][]{Lowi15, SPT15}.  However, these endeavors continue to
fall short of providing a successful account of the full range of
galaxy properties and scaling relations \citep[\eg][]{SM12, Wein15}.
Here we pursue a complementary strategy of using more limited, and
less computationally expensive, simulations to attempt to understand
pieces of the dynamics of galaxy formation in order to develop insight
into the mechanisms that have led to the present-day structure of disk
galaxies.

One of the principal unsolved challenges presented by disk galaxies is
the frequency of bars.  When observed in the near-IR, very roughly one
third are strongly barred, a further one third are weakly barred,
while the remainder appear not to contain any bar at all
\citep{Eskr00, MJ07, Rees07, Buta15}.  Several explanations for the
presence or absence of a bar have been proposed, though none gives a
convincing explanation for these proportions.

\begin{itemize}
\item \citet{OP73}, \citet{ELN82}, and \citet{CST95} have tried to tie
  the existence of bars in disks to halo fraction, but this idea
  collides with both theoretical arguments and empirical data.
  Crucially, these stability criteria were derived for rigid halos and
  are inadequate when halos are composed of responsive particles, as
  reviewed below.  In addition, \citet{Toom81}, who elucidated the
  mechanism of the linear bar instability, argued that bar formation
  can be prevented by inhibiting feedback through the center
  \citep[see also][]{BT08, Sell13}.  On the observational side, the
  great majority of spiral galaxies, both with and without bars, have
  two spiral arms \citep{Davi12, Hart16}, which is an indicator of a
  heavy disk \citep{SC84, ABP87}.  Also, were barred galaxies to have
  heavier disk mass fractions, there would be a systematic offset
  between barred and unbarred galaxies in the Tully-Fisher relation,
  which is not observed \citep[][see also Bosma 1996]{MF96, Cour03}.

\item \citet{Bara09} found an anti-correlation between bulge
  half-light and bar frequency which is weak evidence in support of
  Toomre's (1981) suggestion; feedback through the disk center is
  prevented by an inner Lindblad resonance that should be present when
  the disk hosts a massive bulge.  However, bulge prominence is not a
  reliable predictor for the existence of a bar, as many early-type
  barred galaxies have dense bulges, and at least some late-type
  unbarred systems have insignificant bulges and gently rising
  rotation curves.  While Toomre's linear-theory prediction has been
  demonstrated to work in some careful simulations \citep{SM99, SE01},
  bars still form in other cases \citep{Sell89, PBJ16}, possibly
  through non-linear effects.  It is also possible that dense centers
  have been built up subsequently to the formation of a bar, perhaps
  because of inflows driven by the bar in the disk.

\item Bars are inhibited in disks having large degrees of random
  motion \citep{AS86}.  But since spiral activity is even more
  strongly suppressed, random motion cannot be the explanation for
  unbarred disks that display spiral patterns.

\item \citet{PF90} argued that bars could be destroyed by central mass
  concentrations.  Studies by \citet{NSH96}, \citet{SS04},
  \citet{ALD05} and others have shown that the large central masses
  are required to destroy bars entirely, perhaps $\ga 5\%$ of the disk
  mass.  \citet{KK04} and \citet{Korm13} argue both that such central
  masses can be built up by bar-driven gas inflow and that the
  consequent destruction of the bar creates a pseudo-bulge.  However,
  unbarred galaxies may have neither classical nor pseudo bulges;
  Table~1 of \citet{Korm10} lists several well-studied cases.

\item \citet{Cheu13} find, from Galaxy Zoo data, a greater bar
  fraction in galaxies having lower star-formation rate and argue that
  bar frequency is anti-correlated with gas fraction, as was also
  reported by \citet{Mast12}.  Both these papers use visual ($r$-band)
  images in which the overall bar fraction is $\la 30\%$
  \citep{Mast11}; a substantially larger bar fraction is found in NIR
  images \citep[\eg][]{Eskr00, MJ07}, which suggests that dust and
  star formation can mask bars in visual images of gas-rich systems
  \citep[see][for a remarkable case]{BW91}.  Even with this caveat,
  \citet{Mast12} gave a barred fraction of $\sim 40\%$ in galaxies
  with low \ion{H}{1} mass fractions and 32\% in galaxies with no
  detectable neutral hydrogen, implying that unbarred, gas-poor
  galaxies are not uncommon.

\item Finally, bars could be tidally triggered \citep{Byrd86, Nogu87,
  Geri90, Salo91, Roma08, Gajd16}.  Several studies have examined bar
  fractions as a function of galaxy environments \citep{Elme90, Li09,
    Ague09, Mari12, Lin14}, although the largest sample \citep{Skib12}
  confusingly found that the bar fraction was greatest when galaxies
  were a few diameters apart and decreased at both larger and smaller
  separations!

\end{itemize}

We do not attempt to offer a new explanation for the observed bar
frequency here.  Instead our purpose is to draw renewed attention to
the issue and to show that apparently quite reasonable
stellar-dynamical models of growing disks are pathologically unstable
to bar-formation.  We do not include an explicit dissipative gas
component in our simulations, which some argue is crucial, but our
conclusions at least remain relevant to gas-poor disks where the
problem persists.

As noted above, disks in live halos form bars far more readily than in
the equivalent rigid halo \citep{Atha02}.  \citet{Sell16} demonstrated
that the increased growth rate of the instability resulted from a
torque from the halo acting on the disk even at very low amplitude.
The instability in his toy model simulations that lacked bulges
remained a standing wave with reflection off the center that differed
from the mechanism described by \citet{Toom81} only in that the bar is
excited still more vigorously due to angular momentum loss to the
halo.

\citet{SN13} also reported that bar growth was strongly influenced by
the amount and sense of angular momentum in their live halo models.
They found that halos rotating in the same sense as the disk caused
bars to grow more vigorously and to reach larger size, whereas
counter-rotating halos had the opposite effect.  A related result was
reported by \citet{Sell16}, who showed that the strength of the
disk-halo coupling depended on the shape of the velocity ellipsoid in
non-rotating halos.

These authors, as well as \citet{Atha08}, stressed that the increased
tendency for disks to form bars when embedded in responsive halos
invalidates the disk stability criteria that were derived for rigid
halos \citep{OP73, ELN82}; yet these obsolete stability criteria are
still widely invoked.

The vast majority of barred galaxy simulations have begun with
unstable axisymmetric disks that quickly form bars.  Such simulations
have elucidated the instability, as noted above, and have revealed the
strong secular growth of the bar \citep{AM02, MVSH}.  But the
implication of this body of work for the formation and evolution of
real galaxies is less clear, since the initial unstable models are
most unlikely to have arisen in nature.  We therefore study the
evolution of growing disks, in which the bar instability may assert
itself as the disk is being assembled.

In Paper I \citep{BS15}, we showed that non-uniformities in the
surface density of a galaxy, and consequent features in the rotation
curve, were smoothed out by spiral instabilities.  Using highly
idealized simulations that mimicked a growing disk in rigid halo and
bulge components, we added particles according to rules that were both
somewhat realistic, reflecting inside out disk growth
\citep[\eg][]{MF89, Wang11, Bird13}, and also quite unrealistic.  The rules in
the latter category were attempts to create galaxy models having
unrealistic rotation curves, but the self-consistent dynamics of the
disk always frustrated those efforts.  In that paper, we deliberately
employed rigid mass distributions to represent the halo and bulge in
order to show that the radial rearrangement of mass by spiral activity
in the disk was alone responsible for the smoothing effect.

In this paper, we report a generalization of this study to include
responsive spheroidal components composed of massive particles.  We
expected the evolution of the rotation curve to differ when the bulge
and halo components are made of responsive matter for just two
reasons.  First, a responsive component compresses as the disk mass
rises, due to deepening of the potential well, and second, angular
momentum could be transferred from the disk to the spheroidal
components, although the effects of this transfer on the radial mass
profile are not straightforward.  For example, \citet{Sell03} and
\citet{CVK06} reported additional contraction of the halo in barred
galaxy models; this occured because angular momentum loss from the
inner disk, both to the halo and to the outer disk, concentrated the
disk mass and further deepened the central potential.  This effect
overwhelmed any tendency for the halo density to decrease due to the
work done on the halo by friction from the bar.

We here report that bars formed in most of the live halo models that
matched the stable, rigid bulge/halo models of Paper I.  Our models
therefore present additional examples that illustrate the excitation
of bars by live halos.  Even models having moderately massive, dense
bulges, which would be predicted to be stable by normal mode analysis
\citep{Zang76, Toom81, BT08, Sell13}, formed bars.  We suspect that
they formed through trapping by strong spirals as \citet{Sell89} had
described, but \citet{PBJ16} offer a quite different explanation for
the same behavior in their simulations, and the precise mechanism in
these cases therefore deserves a separate study.

Our original objective was simply to compare the evolution of the mass
distributions and rotation curves in models with live halos with those
in Paper I.  However, bar formation became a major issue that has made
the entire project a much greater challenge.  The formation of bars
not only confuses the interpretation of spiral-driven evolution, but
our prescription, described in \S\ref{sec:addpcs}, of adding fresh
particles to the disk on near circular orbits ceases to be appropriate
when the potential is strongly non-axisymmetric.  Although
non-axisymmetric forces decay rapidly at radii outside a bar, we
report in \S\ref{sec:ebars} that bars grow in size when fresh
particles are added outside the barred part of the disk!  For these
reasons, we have tried, with limited success, to avoid bars in our
models, and those we have selected to report in \S\ref{sec:smoothing}
are some of the few we have run in which the influence of a bar is
minor.  We study the dynamical consequences of responsive halos and
bulges by direct comparisons with models having rigid halos.

\begin{deluxetable*}{ccccccccccccccc}
\tablecolumns{15}
\tablecaption{Simulation information -- physical models}
\tablehead{
& \colhead{Disk} & \colhead{Disk} & \colhead{Initial} & \colhead{Initial} & \colhead{Mass} & \colhead{Mean} & \colhead{Annulus} & \colhead{Bulge} & \colhead{Bulge} & \colhead{Bulge} & \colhead{Halo} & \colhead{Halo} & \colhead{Halo} & \colhead{Halo} \\
\colhead{Run} & \colhead{type} & \colhead{edge} & \colhead{mass}&\colhead{$Q$}&\colhead{added/$\tau_0$}&\colhead{radius}&\colhead{width}& \colhead{type}  & \colhead{mass}  & \colhead{scale} & \colhead{type} & \colhead{mass} & \colhead{scale} & \colhead{$r_{\rm max}$} \\
}
\startdata\\
GR\Ignore{331} & exp           &  4             & 1 & 1.5 & -3.3 & 4 & 4  & none            &                 &                 & Hern      & 36             &  10 & \\
GL\Ignore{373} & exp           &  4             & 1 & 1.5 & -3.3 & 4 & 4  & none            &                 &                 & Hern      & 36             &  10 & 30 \\
GL$^\prime$\Ignore{404}   & exp &  4             & 1  & 1.5 & none & &      & none            &                 &                 & Hern      & 36             &  10 & 30  \\
%ZL\Ignore{368} & exp           &  7.5            & 0.01 & 0.9 & -3.5 & text & text & none    &                 &                 & Isot     & 0.72            &  2 & 80 \\
%YL\Ignore{382} & exp           &  7.5             & 0.01 & 0.9 & -3.5 & text & text & none    &                 &                 & Isot     & 0.72             &  2 & 80 \\
CR\Ignore{305} & KT            &  5             & 0.1  & 1.5   &-3.69 & 6 & 2 & Hern     &  0.9        & 0.2        & Isot     & 14.7         &  30 & \\
CR             &               &                &      &      & 8 & 2 \\
CR             &               &                &      &     & 10 & 2 \\
CR             &               &                &      &     & 12 & 2 \\
CL\Ignore{448} & KT            &  5             & 0.1  & 1.5   &-3.1 & 6 & 2 & Hern         &  0.9            & 0.5             & Isot     & 14.7           &  30 & 150 \\
BL\Ignore{449} & KT            &  5             & 0.667 & 1.5  &-2.69 & 12 & 4 & Hern          &  0.9            & 0.5             & Isot     & 14.7           &  30 & 150 \\
B$^\prime$L\Ignore{470} & KT$^*$        &  5             & 0.2 & 1.5  &-2.62 & 7 & 2 & Hern          &  0.9            & 0.5             & Isot     & 14.7           &  30 & 150 \\
G$^\prime$L\Ignore{445}   & exp &  4             & 1 & 1.5   &-2.3 & 6 & 2 & none           &                 &                 & Hern      & 36             &  10 & 50  \\
G$^\prime$R\Ignore{474}   & exp &  4             & 1 & 1.5   &-2.3 & 6 & 2 & none            &                 &                 & Hern      & 36             &  10 & 50 \\

 \enddata

 \tablecomments{Column 1: Letter identification for the simulation,
   followed by either R for rigid, or L for live halo. Column 2: Type
   of initial disk; ``exp'' is exponential (eq.~\ref{eq:exp}) and
   ``KT'' is the Kuzmin-Toomre disk (eq.~\ref{eq:kt}) and the asterisk
   indicates a modification that is described in the text.  Column 3:
   Initial outer radius of the disk in units of $a$.  Column 4:
   Initial disk mass $q_d$.  Column 5: Initial $Q$.  Column 6: Log of
   fraction of unit mass $M$ added per dynamical time $\tau_0$.
   Column 7: Mean radius of added particles.  Column 8: Width of the
   annulus.  Column 9: Type of bulge, if one is present, ``Hern'' is
   the Hernquist sphere (eq.~\ref{eq:hern}).  Column 10: Mass of the
   bulge component $q_b$.  Column 11: Radial scale of the bulge
   component $b/a$.  Column 12: Type of halo, ``Hern'' is the
   Hernquist sphere (eq.~\ref{eq:hern}), ``Isot'' is the cored
   isothermal sphere (eq.~\ref{eq:isot}).  Column 13: Mass of the halo
   component $q_b$ or $q_c$.  Column 14: Radial scale of the halo
   component $b/a$ or $c/a$. Column 15: Outer edge of halo, $r_{\rm
     max}/a$}.
 \label{tab:lruns}
\end{deluxetable*}

%\newpage
%-----------------------------------------------------------
\section{Live halos} \label{sec:lhalos}
%-----------------------------------------------------------

Here we present results from simulations that resemble those in Paper
I, but in which the halo and bulge are represented by live populations
of particles.  We determine the initial equilibria of these components
as follows.

%-----------------------------------------------------------
\subsection{Model set up} \label{sec:setup}
%-----------------------------------------------------------

Our bulge and halo components are created from spherical mass models
having known distribution functions (DFs) when in isolation.  We then
compute the equilibrium DF of the spherical components in the
composite disk, halo, and bulge model by recognizing that adiabatic
compression leaves the DF unchanged when expressed as a function of
the actions \citep[\cf][]{Youn80}.  In the case of spherical
components, only two actions need be considered: the radial action and
the total angular momentum.  Details of the iterative solution for a
single equilibrium spheroid with an added disk were given in
\citet{SM05} and its generalization to include multiple spherical
components is described here in the Appendix.

\def\pad{\phantom{0}}
\begin{deluxetable*}{rrlcclrcc}
\tablecolumns{9}
\tablecaption{Simulation Information -- numerical parameters}
\tablehead{
 & \colhead{Cylindrical} &  &\colhead{Softening} & \colhead{Spherical} & \colhead{Time} & \colhead{End} & \colhead{Initial}   & \colhead{Final} \\
\colhead{Run} & \colhead{polar grid}  & \colhead{$\delta z$}& \colhead{length}&  \colhead{grid} & \colhead{step} & \colhead{time} & \colhead{particles } & \colhead{particles} 
}
\startdata\\
GR\Ignore{331}  & $106\times128\times135$ & 0.1 & 0.2\pad &  & 1/100 & 8000  & 6 & 6.7  \\
GL\Ignore{373}  & $106\times128\times125$ & 0.02 & 0.2\pad & \pad301 & 1/100 & 8640  & 6 & 6.7  \\
GL$^\prime$\Ignore{404} & $100\times128\times125$ & 0.02 & 0.2\pad & \pad301 & 1/100 & 1500  & 6 & 6 \\
%ZL\Ignore{368} & $160\times256\times125$ & 0.02 & 0.05 & 1001 & 1/64 & 4000   & 4 & 6 \\
%YL\Ignore{382} & $180\times256\times125$ & 0.02 & 0.05 & 1001 & 1/64 & 4000 & 4 & 6   \\
CR\Ignore{305} & $128\times128\times135$ & 0.1 & 0.2\pad &  & 1/100 & 50000  & 5 & 6.71  \\
CL\Ignore{448} & $224\times256\times125$ & 0.1 & 0.2\pad & 1001 & 1/200 & 2500  & 6 & 7.14  \\
BL\Ignore{449} & $224\times256\times125$ & 0.1 & 0.2\pad & 1001 & 1/640 & 2250  & 6 & 6.83  \\
B$^\prime$L\Ignore{470} & $100\times128\times125$ & 0.1 & 0.2\pad & \pad501 & 1/400 & 1100  & 6 & 7.15  \\
G$^\prime$L\Ignore{445}  & $100\times128\times125$ & 0.02 & 0.2\pad & \pad301 & 1/100 & 2000  & 6 & 6.8 \\
G$^\prime$R\Ignore{474}  & $100\times128\times125$ & 0.02 & 0.2\pad & \pad301 & 1/100 & 2000  & 6 & 6.8 \\

 \enddata

 \tablecomments{Column 1: Identification, as column 1 of
   Table~\ref{tab:lruns}. Column 2: number of rings, spokes and planes
   in the cylindrical polar grid.  Column 3: Vertical distance between
   grid planes.  Column 4: Gravity softening length, Newtonian forces
   apply at distances greater than twice this length.  Column 5: The
   number of shells in the spherical grid.  Column 6: 
   The basic time step in units of $\tau_0$.  Column 7: The duration of
   the simulation.  Column 8: Log of initial number of disk
   particles. Column 9: Log of final number of disk particles.}
 \label{tab:nruns}
\end{deluxetable*}

The models we present are listed in Table~\ref{tab:lruns}, which are
selected from the $\sim 100$ simulations we have run in the course of
this project.  Almost every one of our models having responsive halos
formed strong bars.  We have opted to present mostly those in which
bar formation was sufficiently delayed that the smoothing effect of
the spirals could be established.

We employed the \citet{Hern90} sphere for the bulge in some models and
for the halo in others.  It has the density profile
\begin{equation}
\rho_b(r) = {Mq_b  \over 2 \pi b^3}
\left({r \over b}\right)^{-1}\left(1+{r \over b}\right)^{-3},
\label{eq:hern}
\end{equation}
where $r$ is the spherical radius, $Mq_b$ is the component mass, and
$b$ is a length scale.  The equilibrium isotropic DF was also given by
\citet{Hern90}.  In other simulations, we used a cored isothermal
density profile for the halo, for which we determined the isotropic DF
by Eddington inversion \citep{BT08}.  The density profile is
\begin{equation}
\rho_c(r) = {Mq_c \over 4\pi c^3}{3+(r/c)^2 \over [1+(r/c)^2]^2} \ ,
\label{eq:isot}
\end{equation}
where $c$ is the core radius and $Mq_c$ is a mass such that the
asymptotic circular speed $V_\infty = (GMq_c/c)^{1/2}$.  As both these
halos have infinite extent, we eliminate all particles with enough
energy to pass beyond $r_{\rm max}$, which results in the density
decreasing smoothly to zero at that radius.

We employ two types of disk.  The usual exponential
\begin{equation}
\Sigma(R) = {Mq_d \over 2\pi a_e^2} \exp^{-R/a_e},
\label{eq:exp}
\end{equation}
and the Kuzmin-Toomre disk \citep[model 1 of][]{Toom63}
\begin{equation}
\Sigma(R) = {Mq_d \over 2\pi a_k^2} \left[1 + \left({R \over a_k}\right)^2\right]^{-3/2}.
\label{eq:kt}
\end{equation}
In these expressions, $Mq_d$ is the disk mass, and $a_e$ and $a_k$ are
length scales.  The surface density is tapered smoothly, over half a
length scale, to zero at the truncation radius given in
Table~\ref{tab:lruns}.

Having determined the equilibrium DF of each spherical component in
the composite rigid disk, bulge and halo potential, we used it to
select particles as described in the appendix of \citet{DS00}.

We also realize the disk with particles on near circular orbits in the
total potential.  The initial radial velocity spread is set to create
the desired $Q$ value as
\begin{equation}
\sigma_R = Q(R) \sigma_{R, \rm crit} \quad {\rm where} \quad
\sigma_{R, \rm crit} = {3.36 G \Sigma \over \kappa}.
\end{equation}
Here $\kappa(R)$ is the local epicyclic frequency.  Since the velocity
dispersions are a small fraction of the circular speed, $v_c$, in our
initial sub-maximal disks, the (smaller) azimuthal dispersion can be
determined using the epicyclic relation
\begin{equation}
\sigma_\phi = {\kappa \over 2\Omega}\sigma_R,
\end{equation}
where $\Omega(R) = v_c/R$.  We then use the Jeans equation
\citep[][eq.~4.227]{BT08} to estimate the mean orbital speed
$\overline{v_\phi}$
\begin{equation}
\overline{v_\phi}\,^2 = v_c^2 - \sigma_\phi^2 + \sigma_R^2 \left[1 + {d
  \ln(\Sigma\sigma_R^2) \over d\ln R} \right].
\end{equation}
Disk particles are spread vertically with a Gaussian profile and the
$z$ velocities are determined by integrating the vertical 1D Jeans
equation \citep[][eq.~4222b]{BT08} for a slab
\begin{equation}
\sigma_z^2(R,z) = {1 \over \rho(R,z)} \int_z^\infty \rho(R,z^\prime)
      {\partial \Phi \over \partial z^\prime} \; dz^\prime,
\end{equation}
which neglects any radial gradients.  The vertical gradient of the
total potential $\Phi$ is determined from the particles, as well as
any additional mass components.  Because the initial disks have quite
low mass, this standard procedure results in disks that are excellent
equilibria, as are the spherical components.

We adopt $M$ as our mass unit, and the disk scale $a$ as the unit of
length.  Thus our unit of velocity is $V_0 = (GM/a)^{1/2}$ and time
unit, or dynamical time, is $\tau_0 = a / V_0 = (a^3/G M)^{1/2}$.
Henceforth, we use units such that $G=M=a=1$.  A suitable scaling to
physical units for most of our models is to choose $a = 0.5\;$kpc and
$\tau_0 = 1.5\;$Myr, which yields $M \simeq 1.2 \times
10^{10}\;$M$_\odot$ and $V_0 \simeq 326\;$km/s, but other scalings
would be more appropriate for some of our models.

%\newpage
%-----------------------------------------------------------
\subsection{Disk growth} \label{sec:addpcs}
%-----------------------------------------------------------

As in Paper I and in \citet{SM99}, we added particles to the disk
mid-plane, giving each zero radial and vertical speeds, with the
azimuthal speed $v_a = (-Ra_R)^{1/2}$, where $a_R$ is the radial
component of the potential gradient at that point and time.  This
procedure loosely mimics the growth of stellar disks, since stars form
from gas that has settled into approximate centrifugal balance.  The
technique was first devised by \citet{SC84}, and has been adopted
recently in other work \citep{AS15, ABS16}.

This rule gives reasonable initial velocities to the particles only
while the potential remains approximately axisymmetric.  Since our
science goal was to study the consequences of spiral activity, we stop
most of our simulations once a dominating bar develops, and our simple
rule remains adequate until this moment.  Exceptions are presented in
\S\ref{sec:bars}.

Not only is our rule computationally more efficient than modeling the
hydrodynamics of cooling halo gas, star formation and feedback, but it
both gives us far more direct control over the radial location of the
added mass and the dynamical consequences are not complicated by
``gastrophysics''.  We exploited that freedom to the full in Paper I,
where we placed particles on circular orbits over some adopted narrow
radial range.

In all the models reported here, the new particles were added at a
constant rate in a narrow annulus with the radial range given in
Table~\ref{tab:lruns}.  Our objective is not to match any particular
expectation of the angular momentum distribution of the material that
makes up the disk, but to confirm for live halos what we already
showed in Paper I for rigid halos, that the profiles of the evolved
disks are insensitive to the distribution of angular momentum added to
the disk.

%----------------------------------------------------------------------
\begin{figure}
\begin{center}
\includegraphics[width=.9\hsize,angle=0]{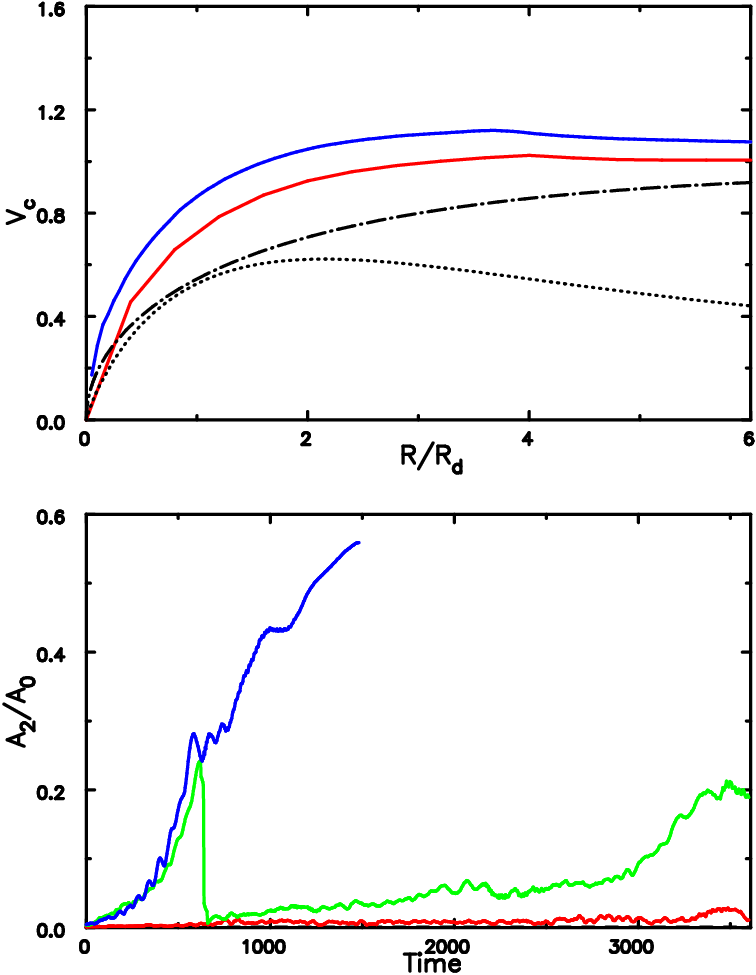}
\end{center}
\caption{Upper panel: the rotation curves of three models without a
  bulge.  The dotted and dot-dash lines show the disk and the rigid
  halo contributions, while the colored lines show the total circular
  speed: red for the rigid halo case GR (model G of Paper I), and blue
  for the live halo GL used here, which employed the same initial halo
  but which was compressed by the growth of the disk.  The lower
  panel: the time evolution of the bar amplitude.  The red line is for
  model GR, which had a growing disk in a rigid halo, the green line
  shows the result for GL, a growing the disk in a live halo, with the
  bar being destroyed by fiat at $t=640$, and the blue line shows
  GL$^\prime$ the same initial model as GL but with no accretion,
  showing secular bar growth.}
% rigid halo 331, live halo with accretion 373, live halo, no accretion 404
\label{fig:barg}
\end{figure}
%----------------------------------------------------------------------

%-----------------------------------------------------------
\subsection{Numerical method} \label{sec:code}
%-----------------------------------------------------------

We use the hybrid $N$-body code described by \citet[][Appendix
  B]{Sell03}, in which the gravitational field is computed with the
aid of two grids: a 3D cylindrical polar grid for the disk component,
and a spherical grid for the bulge/halo components with a surface
harmonic expansion on each grid shell up to order $l_{\rm max}=4$.  A
full description of our numerical methods is given in the on-line
manual \citep{Sell14}.

We measure the amplitude of non-axisymmetric features by summing
\begin{equation}
A_m(t) = \left| \sum_j \mu_je^{im\phi_j} \right|,
\label{eq:defamp}
\end{equation}
for disk particles only, where $\mu_j$ is the mass and $\phi_j(t)$ the
cylindrical polar angle of the $j$-th particle at time $t$.  We define
the bar amplitude as $A_2/A_0$.  We estimate the length of a bar in a
simulation as described in the Appendix of \citet{SD06}.  Briefly, it
is the average of two separate estimates: (a) the radius at which the
$m=2$ amplitude decreases to half its peak value and (b) the radius at
which the phase of the same sectoral harmonic differs by more than
$20^\circ$ from the mass-weighted average near the peak.

We employed one million particles for the halo, and another million of
the bulge, if present.  The initial and final numbers of particles in
the disk are given in Table~\ref{tab:nruns}, which also lists the
other numerical parameters used for each run.  As usual, we checked
that outcomes were little changed by reasonable variations of particle
number, time step, and grid resolution.  Michael Aumer (private
communication) kindly confirmed that responsive halos also lead to
higher growth rates for bars when a tree code is used instead.

%\newpage
%-----------------------------------------------------------
\section{Bar formation} \label{sec:bars}
%-----------------------------------------------------------
\subsection{No bulge} \label{sec:barg}
%-----------------------------------------------------------

The upper panel of Figure~\ref{fig:barg} illustrates the initial
rotation curves of three models that consist of a disk and halo only.
The red curve is from model GR (the same as model G of paper I) which
was a submaximal exponential disk in a rigid Hernquist halo, as
indicated by the dotted and dot-dash lines respectively.  The blue
curve is for two initially identical models GL that started with the
same halo as model GR, but which had been compressed by the growth of
the initial disk, and was realized by responsive particles.

The lower panel gives the time evolution of the bar amplitude.
The red line shows that the $m=2$ amplitude remained very low in model
GR, despite the fact that accretion of particles on near-circular
orbits caused the disk mass to grow to 2.8 times its initial mass by
the last moment shown, and to 4.5 times the initial mass by the time
we ended the calculation.  Thus this model appeared to be robustly
stable.

\begin{figure}
\begin{center}
\includegraphics[width=.9\hsize,angle=0]{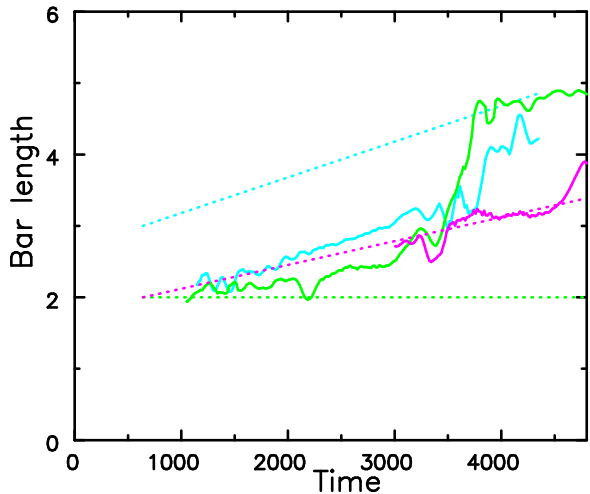}
\end{center}
\caption{The time evolution of the bar semi-major axis in run GL
  (green), and two similar models.  The dotted lines indicate the
  inner radius of the accretion annulus in the model of the
  corresponding color.  The early evolution to $t=640$ in which a bar
  formed and was destroyed, as in Fig.~\ref{fig:barg}, is not shown.}
\label{fig:shwres}
\end{figure}

The behavior in the live halo is illustrated by the other two lines.
Even though halo compression further reduced the effective mass
fraction of the disk, the live halo encouraged rapid bar formation, as
found by others.  The green line shows model GL, in which we added
disk particles by the same rule as in the rigid halo.  When we saw
that a bar had formed, we destroyed it at $t=640$ by randomizing the
azimuths of all the disk particles as we did in Paper I.  This
intervention therefore created a dynamically hot, featureless disk to
which we resumed adding fresh particles.  While the hot disk was
stable, the effect of the additional cool particles was to make the
model become unstable again, and a new bar developed after $t=3000$.
The time evolution of the bar length in model GL is shown by the green
line in Fig~\ref{fig:shwres}, which is described more fully below.

%----------------------------------------------------------------------
%----------------------------------------------------------------------

%----------------------------------------------------------------------
\begin{figure*}
\begin{center}
\includegraphics[width=.6\hsize,angle=270]{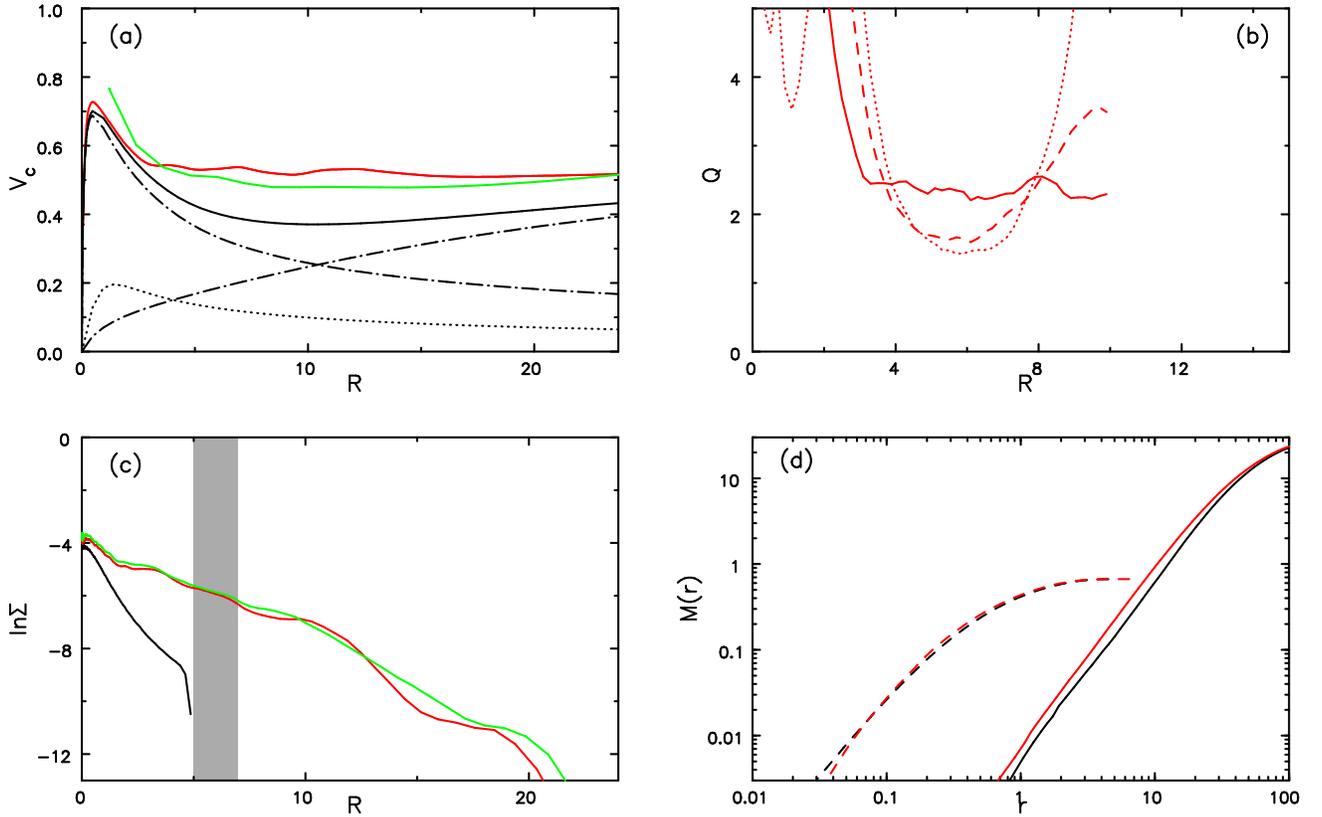}
\end{center}
\caption{(a) The colored curves show the circular speed in run CL
  (red) at $t=1440$ and in run CR (green) at the same accreted mass.
  The initial rotation curve is shown by the black lines; the dashed
  line is the total, with the contributions from the bulge and halo
  are given by the dot-dash lines, and the disk by the dotted line.
  (b) The radial $Q$ profiles in model CL at $t=480$ (dotted), $t=960$
  (dashed), and $t=1440$ (solid).  (c) The initial (black) and final
  (color) surface density profiles, with shading indicating the radial
  range of accretion.  (d) The mass profiles of the bulge (dashed) and
  halo (solid) with the initial profile in black and that at $t=1440$
  in red.}
% The original run C was run 305, the live halo run is 448
\label{fig:figC}
\end{figure*}
%----------------------------------------------------------------------

%-----------------------------------------------------------
\subsection{Large bars} \label{sec:lbars}
%----------------------------------------------------------- 

One of the major differences between the models used here and those
employed by \citet{Sell16} is the much more extensive halo.  The blue
curve in the lower panel of Fig.~\ref{fig:barg} illustrates the bar
amplitude in model GL$^\prime$ that started out with the same disk and
halo as GL, but without the addition of new particles.  By the time we
stopped the calculation, the bar's semi-axis was $a_B \simeq 4a$,
which was equal to the outer radius of the initial disk!  This
continued secular growth of the bar was caused by on-going angular
momentum exchange with the extensive halo; the loss of angular
momentum allowed more disk particles to become trapped into the bar.

Model GL$^\prime$ therefore manifests the same continuous growth of
bar size and strength that has been reported previously \citep{AM02,
  MVSH}.  We have verified that a model set up to be as similar as
possible to the ``massive halo'' (MH) case presented by \citet{AM02}
did indeed evolve with our quite different code in the manner they
reported.  In particular, the bar continued to grow to $t=900$, by
which time it was both long (semi axis $\sim 5a$) and very strong
($A_2/A_0 \sim 0.5$), as the similar case GL$^\prime$ shown by the
blue curve in Fig.~\ref{fig:barg}.

%\newpage
%-----------------------------------------------------------
\subsection{Disk excitation of bars} \label{sec:ebars}
%----------------------------------------------------------- 

While the discussion above has focused on halo excitation of bars,
we here report that the cool outer disk also shares the responsibility
for large strong bars.  In particular, it seemed possible to us that
the excitation of a new bar in model GL was caused by too small an
inner radius of the accretion annulus, at $R=2$.  The added material
could couple gravitationally to the inner disk in such a way as to
excite a bar.  We therefore tried two further experiments in which
nothing was changed except that we increased the mean radius of the
accretion annulus continuously to larger radii, in the hope that the
later added particles would be too far out to couple to the inner
disk, thereby avoiding a bar.

However, the models still formed strong bars and, if anything, the bar
length seemed to increase with the radius of the accretion annulus, as
shown in Figure~\ref{fig:shwres}!

The results from model GL are displayed in green in
Figure~\ref{fig:shwres}, as are those in Figure~\ref{fig:barg}.  In
this model the inner edge of the accretion annulus was held fixed, and
therefore some particles that were added within the bar, which always
extended into the accretion annulus, were given inappropriate
velocities.  In an attempt to avoid this unrealistic evolution, we
increased the inner radius of the annulus over time as shown by the
dotted lines.  In the first such experiment, indicated by the magenta
lines, a bar still formed and its semi-major axis seemed to track the
inner edge of the annulus, at least for a while.  The behavior in an
additional experiment that began with a larger accretion annulus that
was also increased a little more rapidly is shown by the cyan lines;
once again a bar formed whose size rapidly approached the inner edge
of the accretion annulus.

It seems that additional cool particles in the outer disk were able to
accept angular momentum from the bar in the inner disk, causing it to
grow in strength and length.  Our strategy of shifting this material
to larger radii as the bar grew seemed to enable this happen more
readily, having the opposite of the desired effect!

%----------------------------------------------------------------------
\begin{figure}
\begin{center}
\includegraphics[width=.9\hsize]{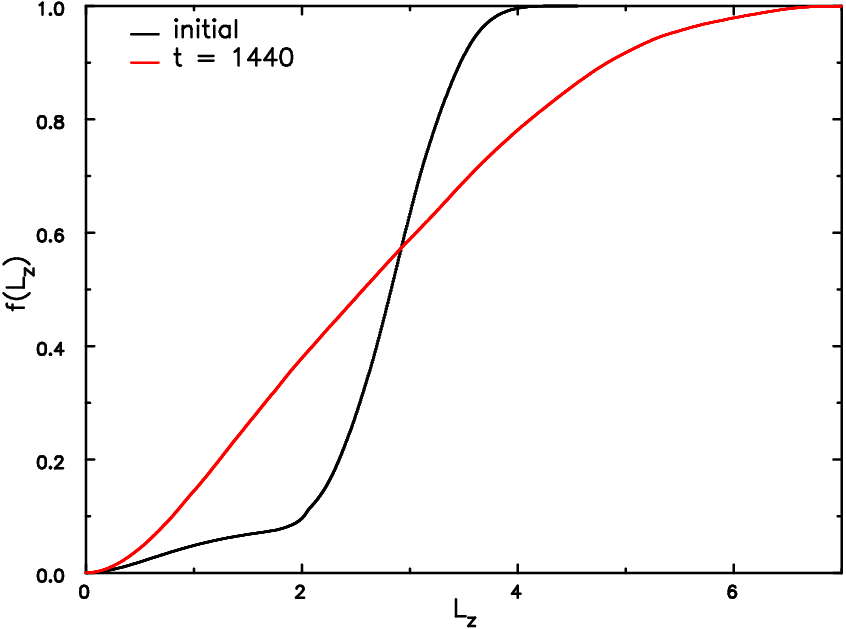}
\end{center}
\caption{Disk mass fraction as a function of angular momentum in run
  CL.  The black curve shows the initial distribution of angular
  momentum while the red curve indicates the final distribution.  Note
  the smoothness of these distributions, which were measured directly
  from the large numbers of disk particles (see
  Table~\ref{tab:nruns}).}
% run 448
\label{fig:nvo}
\end{figure}
%----------------------------------------------------------------------

%----------------------------------------------------------------------
\begin{figure*}
\begin{center}
\includegraphics[width=.6\hsize,angle=270]{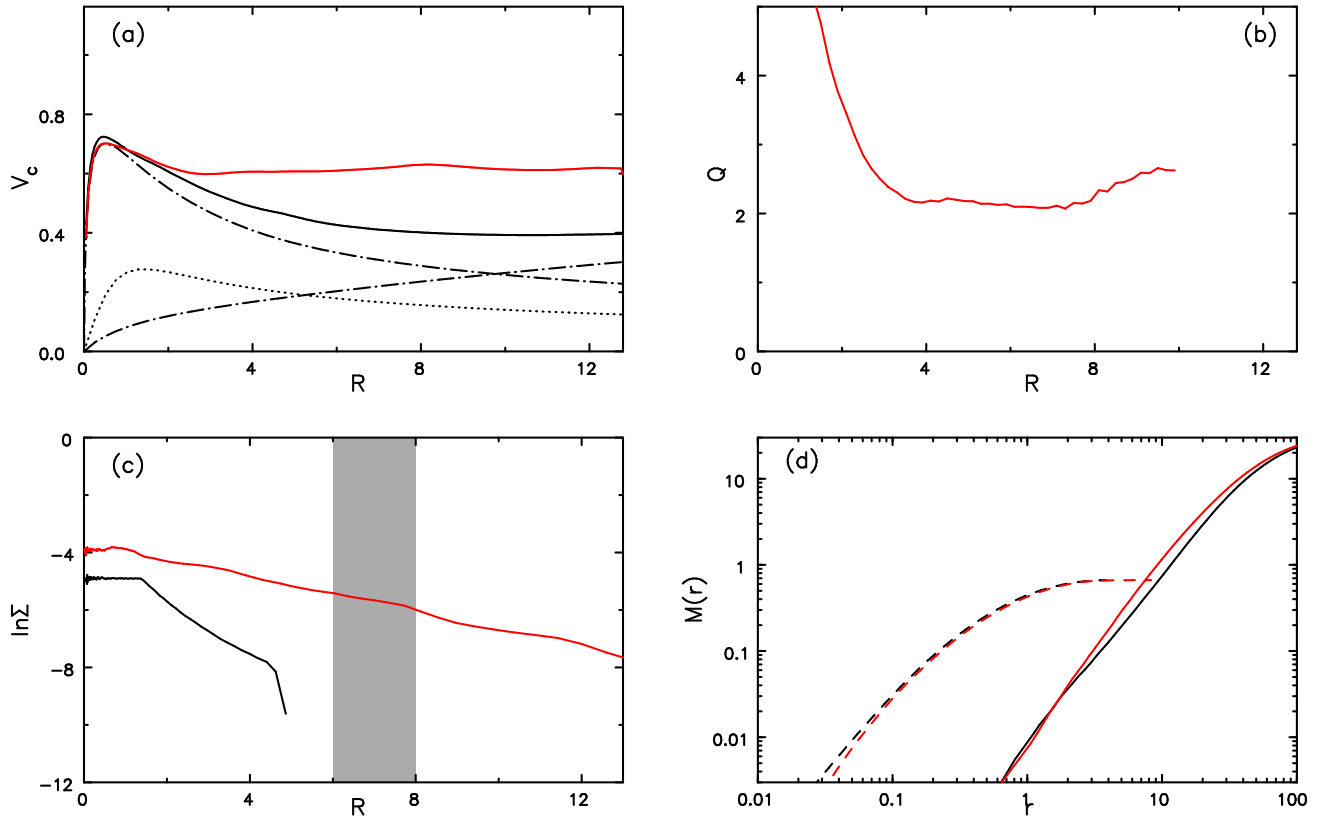}
\end{center}
\caption{As for Fig.~\ref{fig:figC} but for run B$^\prime$L with a live bulge
  and halo only.  The black lines are at $t=0$ and the red lines at
  $t=1000$.}
% live halo run is 470, rigid (472) is too strongly barred to use
\label{fig:figB}
\end{figure*}
%----------------------------------------------------------------------

%-----------------------------------------------------------
\section{Smoothing rotation curves} \label{sec:smoothing}
%-----------------------------------------------------------

In Paper I, we showed that the final mass and angular momentum
distributions in the disk were insensitive to the angular momentum of
the accreted material.  We gave a full description of the smoothing
mechanism in that paper.  Briefly, we showed that spiral modes
rearranged the angular momentum profile of the disk through horseshoe
orbits of stars near the corotation radii of the spirals \citep{SB02},
while the spirals were themselves excited by the same features in the
density profile.

Since we employed rigid mass distributions for the spherical mass
components of those models, our study clearly demonstrated that the
angular momenta of added particles could be diffused by spiral
activity so as to create largely smooth disk density profiles and
rotation curves lacking strong features.  The purpose of this
follow-up study was to determine how this result might be affected by
replacing the rigid spherical components by responsive matter
represented by populations of live particles.

As previously emphasized, most models from our earlier study with
rigid spherical components formed bars when the bulge/halo components
were realized with responsive particles.  We have opted not to pursue
simulations after a strong bar has formed since our science goal is to
study the role of spirals alone.  We here present only simulations
that support a protracted period of spiral activity, although a bar
appeared late in the evolution in most cases, causing us to stop the
simulation soon thereafter.

%-----------------------------------------------------------
\subsection{Massive bulge}
%-----------------------------------------------------------

The red line in Figure~\ref{fig:figC}(a) shows the rotation curve at
$t=1440$ from model CL with live bulge and halo components that
closely resembled the rigid halo case run CR (run C of Paper I); the
result from model CR at the same accreted mass is shown by the green
line.  Both models had a dense Hernquist bulge, a pseudo-isothermal
halo, and an initially very low mass exponential disk.  We added
particles at a constant rate over the radius range $5 < R/a < 7$,
indicated by the shaded region in panel (c), with radii chosen from a
uniform probability distribution.

We added mass more rapidly in CL than in CR, did not shift the
accretion annulus outwards, and we stopped the simulation at the time
shown, when the disk mass had increased 13-fold, because a bar had
developed in CL.  The bar appeared around $t=1300$ as an oval
distortion that extended to $\sim 3a$ and grew in amplitude (to
$A_2/A_0 \sim 0.08$) and length (to $\sim 5a$) by $t=1440$.

In this case, the run was able to manifest the smoothing effect of the
spirals, as bar formation occurred after much mass had been added.
The principal result from Paper I was reinforced: there are no abrupt
features in the rotation curve, despite the limited radial extent of
accretion.  Panel (c) demonstrates that spiral activity spread the
added disk material both inwards and outwards from the annulus in
which it was added in a manner very similar to that we found in the
rigid halo (green line).  Panel (b) shows the $Q$ profile, which was
computed from the rms radial velocities and other appropriate disk
properties at each radius and time, at three representative times.
While the inner disk becomes dynamically hot at an early stage, the
region where particles are added stays cool until the bar forms.

Figure~\ref{fig:nvo} confirms that the distribution of angular
momentum within the disk is substantially rearranged by the spiral
activity.  The black line shows the distribution in the initial disk,
which extended to $L_z \sim 2$ plus the distribution of the added
particles, which were placed on circular orbits outside the initial
disk.  The red line indicates that spiral activity has substantially
smoothed the distribution by the time shown.  As noted in paper I, the
near constancy of $dM_d/dL_z$ is reminiscent of, but not quite
consistent with, the arguments presented by \citet{LH78}.

As expected, halo compression yields a slightly higher circular speed
in CL than in CR for the same disk mass added.  The solid curves in
panel (d) of Fig.~\ref{fig:figC} show the radial mass profiles of the
initial (black) and final (red) halo, indicating the extent of halo
compression by the added disk.  The dashed curves show the same
quantities for the bulge, whose profile is little changed except in
the innermost 1\%.  Very little angular momentum is gained by the halo
from the disk in run CL, with the torque increasing to $dL_z/dt \simeq
10^{-5}$ by the end as mild bar begins to form.  The angular momentum
gained by the bulge is at least an order of magnitude less.

%----------------------------------------------------------------------
\begin{figure*}
\begin{center}
\includegraphics[width=.6\hsize,angle=270]{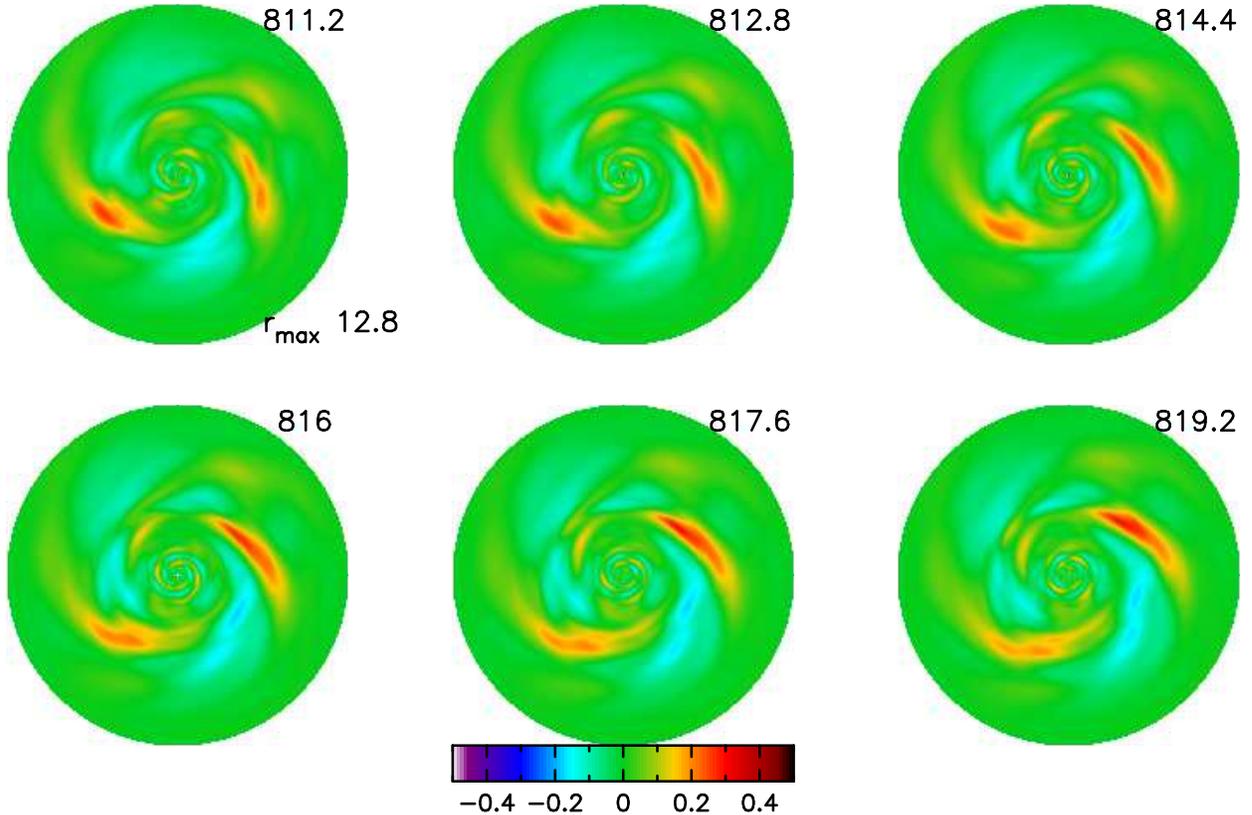}
\end{center}
\caption{A short time sequence late in the evolution of run B$^\prime$L showing
  departures from the mean density at each radius.  Note the tightly
  wrapped spiral in the disk center where the surface density was
  low.}
\label{fig:cntd}
\end{figure*}
%----------------------------------------------------------------------

%----------------------------------------------------------------------
\begin{figure*}
\begin{center}
\includegraphics[width=.6\hsize,angle=270]{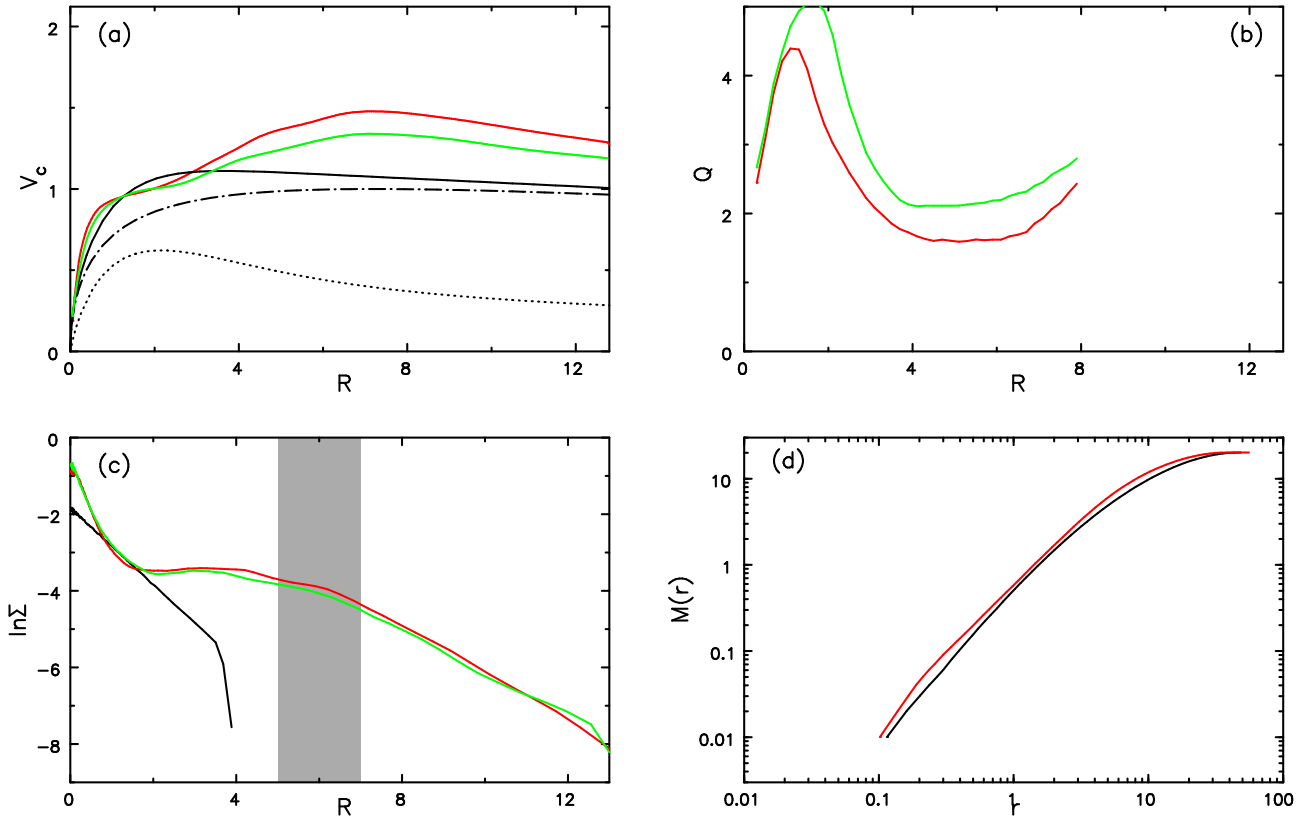}
\end{center}
\caption{As for Fig.~\ref{fig:figC} but for run G$^\prime$L with a
  halo only and no bulge.  The black lines are at $t=0$ and the
  colored lines at $t=2000$: red for G$^\prime$L with the live halo,
  green for G$^\prime$R in which the halo was frozen.}
% live halo run is 445 rigid 474
\label{fig:figG}
\end{figure*}
%----------------------------------------------------------------------

%\newpage
%-----------------------------------------------------------
\subsection{Moderate bulge}
%-----------------------------------------------------------

Run B from paper I had a less massive bulge, and we obtained a similar
result in a parallel simulation with a live bulge and halo, although
we ended the live halo simulation at $t=8000$, whereas we had
continued the rigid halo model to $t=18000$.  Since the orbit period
at $R=5a$ is $\simeq63$ of our time units, the duration of even the
shorter integration was some 125 disk rotations, which is considerably
longer than the age of a Milky Way-like galaxy.

As for model CL above, we therefore ran other simulations with
substantially higher accretion rates, in order to reach the same final
disk mass within the expected lifetime of a galaxy.  However, this
approach was unsuccessful for this lower bulge mass, because the
models formed strong bars well before the desired final mass was
reached.

One such example was model BL.  As in Paper I, we destroyed the
initial bar at $t=750$, after which we began to add particles far
outside the original disk.  However, a new bar formed around $t=1400$
that extended as far as the accretion annulus and grew in amplitude
until we stopped the run at $t=2250$ when the disk mass had increased
just 5-fold.

Since this behavior prevented us from observing the smoothing effect
of spiral activity, we had to develop a strategy that would inhibit
the formation of bars.  After much experimentation, we found the most
effective way to delay the formation of a bar was to reduce
substantially the {\em inner\/} disk surface density in the region of
the bulge.  The evolution run, B$^\prime$L, is shown in
Figure~\ref{fig:figB}.  The initial KT disk profile is given by the
black line in panel (c), with the surface density held constant when
$R\la 1.2$.  We added particles at a constant rate from $t=0$ in the
annulus $6 < R < 8$ such that the disk mass increased 13-fold by
$t=1000$.

As expected, lowering the surface density had the effect of decreasing
the spatial scale of self-gravitating density waves.
Figure~\ref{fig:cntd} illustrates a tightly wrapped $m=2$ spiral in
the inner disk over a short interval late in the evolution of run
B$^\prime$L.  The radial group velocity of spiral waves \citep{Toom69,
  BT08, Sell13} is $v_g = \partial\omega/\partial k$, where $\omega =
m\Omega_p$ is the pattern speed and $k$ the radial wavenumber.
\citet[][their eq.~(6.76), which is missing a ${\rm sgn}(k)$
  factor]{BT08} give the messy expression for the group velocity
derived from the WKB dispersion relation for stellar disks, the
leading factor of which is inversely proportional to $k$.  In this
case therefore, the tighter winding of spiral waves where the surface
density is lowered also likely slows the group velocity.  Delayed
feedback through the center, if it occurs at all, slows the growth
rate of the bar instability, substantially delaying the formation of
the bar.

The smoothing effect of the spirals again dominated the changes to the
circular speed curve (Figure~\ref{fig:figB}a), as well as the surface
density profile (c).  The surface density of the inner disk remained
low for most of the run, but a large bar developed shortly before the
time shown ($t=1000$) that was associated with a significant increase
in the inner surface density.  The effect of halo compression is shown
in (d).

%\newpage
%-----------------------------------------------------------
\subsection{No bulge}
%-----------------------------------------------------------

As shown by the red line in Fig.~\ref{fig:barg}(b), model GR (G of
paper I) did not form a bar even as the disk mass was increased
substantially, but bars formed more readily when the same halo was
made responsive (model GL) even without the addition of disk mass
(model GL$^\prime$).  We could have suppressed bar formation by
increasing the halo mass fraction \citep{MW04, Sell16}, but mass
rearrangements by the small-scale spirals that would develop within
such extreme low mass disks would cause only minor changes to the
shape of the overall rotation curve, which would not achieve our
science goal.

In Paper I, we dealt with bar unstable disks by letting the disk form
a bar and then scrambling the azimuths of the disk particles to erase
the bar and using this hot initial disk as the starting point for mass
accretion.  This growing disk in the rigid halo generally avoided
forming a new bar, at least for long enough to reveal the smoothing
effects of spirals.  However, as shown in Fig.~\ref{fig:barg}, a fresh
bar formed in a live halo when we equally slowly added cool material
to the hot disk made in this manner.

Thus, following \citet{SN13}, we suppressed the supporting response to
the bar mode in the disk by making the halo rotate strongly in the
sense counter to disk rotation.  This strategy is clearly an artifice:
in reality, any halo rotation should have nearly the same sense of
rotation as the disk, because both the halo and the disk material were
subject to the same tidal torques and inflows in the early universe
\citep{Efst79, Bull01, Bett07}.  The disk spins up as it shrinks and
settles through dissipation, while rotation in the more extended
collisionless halo is also expected to be mild and in the same sense
as that of the disk \citep[but see also][]{Dano15}.

We added rotation to the halo in the following manner.  An isotropic
halo has $f(E)$ only, with uniform probability of any allowed value of
the total angular momentum $L$ at that $E$.  We selected particles
with the required distribution of these integrals as described in
\citet{DS00}, and then chose a radius at random from a uniformly
distributed fraction of the radial period.  These choices determine
$v_\perp$ and $v_r$ for the particle, where $v_\perp = L/r$ and $v_r^2
= 2[E-\Phi(r)] - L^2/r^2$ are the components of velocity perpendicular
and parallel to the radius vector.  For an isotropic DF, the direction
of $v_\perp$ is uniformly distributed in $-\pi \leq \psi \leq \pi$,
where $\psi$ is the angle to the $(x,y)$ plane, say.  We can add
arbitrary amounts of net retrograde rotation by biasing the choice of
$\psi$ towards values $|\psi| > \pi/2$.  In our case, we set $\psi =
\pi u^{1/3}$, where $u$ is uniformly distributed over the range $-1 <
u < +1$.

Although a rigid halo of this mass is sufficient to inhibit bar
formation, even the strong retrograde bias of this live halo, model
G$^\prime$L, did not prevent a cool disk from forming a bar.  We
therefore destroyed the bar at $t=640$ by randomizing the azimuths of
all the disk particles, thereby creating a hot, stable disk.  We added
particles to this initial, dynamically hot disk over the radial range
$5 < R/a < 7$ at a higher rate than in model GL and no bar formed
during the subsequent evolution to $t=2000$, when the disk mass was
6.2 times greater than at the outset.  G$^\prime$L was our only model
with a live halo that had no tendency to form a bar; comparison models
with isotropic halos, \eg\ GL in Fig.~\ref{fig:barg}, always formed
fresh bars during the accretion phase.

Figure~\ref{fig:figG} shows the results at $t=2000$; the red lines
show the behavior of the model with a live halo, G$^\prime$L, and the
green lines illustrate the evolution of run G$^\prime$R in which the
live halo was frozen at $t=640$ after the initial formation and
destruction of the bar.  Once again, spiral activity has rearranged
the disk radial density profile in almost the same manner for the live
and rigid halos, as shown in panel (c).  The central density spike was
created as the bar formed in the part of the evolution before
particles were added.  The effects of halo compression are evident in
panels (a) and (d); the black line in (d) indicates the halo mass
profile at the start of the simulation, while the red shows the
profile in G$^\prime$L after the disk mass was increased.  The
difference in the red and green lines in (a) is due almost entirely to
the compression of the live halo.  As in model C, above, less than 2\%
of the final disk angular momentum was given up to the halo over the
interval $960 < t < 2000$.

%--------------------------------
\section{Conclusions} \label{sec:conc}
%----------------------------------

The results presented in the previous section have met our original
science objective of showing that a responsive halo does not alter the
conclusion from Paper I: the disk density profile and rotation curve
of the composite system are naturally smoothed by internal dynamical
evolution as a disk grows in mass.  It is interesting that the end
condition seems to be characterized by a uniform distribution of disk
mass with specific angular momentum (Fig.~\ref{fig:nvo}).  Comparison
between models with rigid and responsive halos reveals that halo
compression plays a more minor role than the spiral activity, which we
identified in Paper I as the primary cause of smoothing.  We also
found that a disk supporting spiral activity does exert a torque on
the halo but, unless a bar develops, only a tiny fraction of the total
disk angular momentum is given up to the halo.

However, these conclusions were reached only after we were able to
overcome the very pronounced tendency for a disk in a live halo to
form a bar where a similar disk in a rigid halo did not.  This
increased tendency for bar formation had been reported by others
\citep{Atha02, SN13}, and was studied in greater depth by
\citet{Sell16} who focused on highly idealized models with halos that
were not at all extended.

Since we wished to examine the effects of spirals without a dominating
bar, we struggled to create initial models that were not strongly
unstable when embedded in responsive halos and bulges.  Unless the
bulge was extremely massive (Fig.~\ref{fig:figC}), our simulations
with dense bulges quickly formed bars, despite the prediction by
\citet{Toom81} that a dense center should prevent feedback and
therefore be stabilizing.

We were able to at least delay the formation of a bar in models with
moderate bulges by reducing the central density of the disk
(Fig.~\ref{fig:figB}).  We also adopted the strategy from Paper I of
allowing a bar to form and then destroying it, by randomizing the
azimuths of the disk particles, thereby creating a dynamically hot
disk (Fig.~\ref{fig:barg}).  However, both these strategies provided
only temporary respite from bar formation as we increased the disk
mass by adding fresh, dynamically cool particles.  We were able to
weaken the coupling of the disk to the halo by making the two
components counter-rotate, an unrealistic strategy that appeared to be
the only way to prevent the eventual formation of bar.

We also report (\S\ref{sec:lbars}), as was also found previously by
\citet{AM02} and \citet{MVSH}, that bars in unstable models with
extended halos become quite unrealistically large and strong
\citep[\eg][]{Erwi05}.  While the extraordinary secular growth of bars
manifested by such models is interesting, they start from cool,
axisymmetric disks that are strongly unstable -- \ie\ from initial
conditions that could not arise in nature.

The dominance of bar formation in our models was indeed remarkable.
Almost every live-halo model presented here formed a bar eventually, a
result that is apparently inconsistent with the existence of unbarred
galaxies in nature, as reviewed in the Introduction.

\citet{Cheu13} and others \citep[\eg][]{BCS05} have claimed that bars
can be avoided in simlations having large gas fractions.  However,
\citet{ABS16} reported simulations that included both dissipative gas
and heavy particles to represent giant molecular clouds, but found
that scattering of star particles by the heavy particles, slowed but
did not inhibit entirely, the development of bars in their models.
Many simulations of galaxy formation in a cosmic context have included
rules to mimic baryonic physics to the most realistic extent possible,
and some \citep[\eg][]{Stin13, AWN14, Snyd15} have reported Milky
Way-like galaxy models that do not possess bars.  The challenge will
be to show not only that these simulations can reproduce the observed
bar fraction, but also that the disks in such models are realistically
massive, dynamically cool, and thin so that they can also support the
generally-observed two-armed spiral patterns.

Since we have employed only collisionless particles, and do not
include the physics of gas, our findings may perhaps apply to gas-poor
galaxies only.  However, a large fraction of such galaxies appear to
be unbarred \citep{Mast12}, and the low gas fractions imply that bars
would be harder to mask with star formation and dust \citep{BB01}.
Note that while large gas fractions are expected in the early stages
of galaxy formation, when we destroyed the early bars, fresh bars
formed as we modeled the continued gradual growth of galaxy disks in a
physically reasonable manner.  Thus the problem is also one of late
evolution, and not just of the early stages.

%--------------------------------
\acknowledgments 
%----------------------------------

We thank the referee for a valuable report that greatly helped us to
clarify the paper.  This work was supported by NSF grants AST/110897
and AST/12117937.

% a kluge to create some white space before the start of the appendix
\section*{\phantom{Appendix}}
%-----------------------------------------------------------
\section*{Appendix}
%----------------------------------------------------------- 

\citet{SM05} described how an isolated equilibrium spherical model
could be compressed to include a second rigid mass component by
assuming that the latter was added adiabatically.  Their algorithm
used the conservation of both radial action and angular momentum, as
first set out by \citet{Youn80}, which therefore takes into account
the extra resistance to compression caused by radial pressure that is
ignored in algorithms that conserve angular momentum only
\citep[\eg][]{BFFP86}.  \citet{SM05} also demonstrated that
asphericity in the rigid mass had negligible effect, even when the
rigid component was a disk, indicating that the change to the halo is
very well determined only by the rigid mass enclosed within a
spherical radius.  We take the finite thickness of the disk into
account when computing this mass, which spreads out the disk mass
significantly only at radii that are within a couple of disk scale
heights of the center.  This algorithm yields the DF for the
equilibrium halo in the composite disk+halo potential, from which halo
particles can be selected.  The final step of the set-up procedure is
to realize the disk with particles.

Here we extend the algorithm to a three component system, in order to
be able to create equilibrium bulges and halos in the presence of a
disk.  As for the single spherical component, the working assumption
is that the DFs of the two spherical components, when expressed as
functions of the actions, do not change as the combined model is
assembled adiabatically.  We therefore select models for the halo and
bulge whose DFs when each is in isolation are known, or can be
determined by Eddington inversion \citep{BT08}.

The first step is to construct 2D tables for the specific energy, $E$,
as functions of the two actions: specific angular momentum, $L$, and
radial action, $J_r$, in the potential well of each isolated spherical
component.  These tables are needed so that the DF, which is usually
expressed as a function of the classical integrals, $f(E,L)$, can
be determined from known values of the actions.

The solution for the equilibrium of the composite system proceeds
iteratively.  Using the total spherical potential of the composite
system at each iteration, we compute an additional 2D table of
$J_r(E^\prime,L)$, where $E^\prime$ is a possible energy.  To evaluate
the DF for either component for some given values of $(E^\prime,L)$,
we use our tables first to map $(E^\prime,L) \rightarrow (J_r,L)$ and
then $(J_r,L) \rightarrow (E,L)$, and set $f(E^\prime,L)=f(E,L)$.  We
can therefore integrate the DF of both spherical components over all
accessible values of $E^\prime$ and $L$ to construct revised estimates
of their densities.  The new density estimates yield an improved
estimate of the combined potential, with the inclusion of the disk
monopole term, from which an improved table of $J_r(E^\prime,L)$ may
be computed.  The procedure generally converges to an acceptable level
in 10-15 iterations.

This procedure can be generalized to determine equilibrium DFs for
separate components in any composite system, with or without rigid
potentials.


\begin{thebibliography}{}

\def\aap{A\&A}
\def\aj{AJ}
\def\apj{ApJ}
\def\apjl{ApJL}
\def\apjs{ApJS}
\def\apss{Ap.\ Sp.\ Sci.}
\def\araa{ARAA}
\def\jcop{J. Comp.\ Phys.}
\def\mnras{MNRAS}
\def\newa{New. Astron.}
\def\PhD{PhD.\ thesis}
\def\nat{Nature}
\def\pf{{\it Phys.\ Fluids}}
\def\PhD{{\it PhD thesis}}
\def\phya{{\it Physica\/} A}
\def\pnas{Proc.\ Nat.\ Acad.\ Sci.}
\def\raa{{\it Res A\&A}}
\def\rpp{Rep.\ Prog.\ Phys.}
\def\Omit#1{{ \etal}}

\bibitem[\protect\citeauthoryear{Aguerri \etal}{2009}]{Ague09}
Aguerri, J. A. L., M\'endez-Abreu, J. \& Corsini, E. M. 2009, \aap, {\bf 495}, 491

\bibitem[\protect\citeauthoryear{Athanassoula}{2002}]{Atha02}
Athanassoula, E. 2002, \apjl, {\bf 569}, L83

\bibitem[\protect\citeauthoryear{Athanassoula}{2008}]{Atha08}
Athanassoula, E. 2008, \mnras, {\bf 390}, L69

\bibitem[\protect\citeauthoryear{Athanassoula \etal}{1987}]{ABP87}
Athanassoula, E., Bosma, A. \& Papaioannou, S. 1987, \aap, {\bf 179}, 23
	
\bibitem[\protect\citeauthoryear{Athanassoula \etal}{2005}]{ALD05}
Athanassoula, E., Lambert, J. C. \& Dehnen, W. 2005, \mnras, {\bf 363}, 496

\bibitem[\protect\citeauthoryear{Athanassoula \& Misiriotis}{2002}]{AM02}
Athanassoula, E. \& Misiriotis, A. 2002, \mnras, {\bf 330}, 35

\bibitem[\protect\citeauthoryear{Athanassoula \& Sellwood}{1986}]{AS86}
Athanassoula, E. \& Sellwood, J. A. 1986, \mnras, {\bf 221}, 213

\bibitem[\protect\citeauthoryear{Aumer \etal}{2016}]{ABS16}
Aumer, M., Binney, J. \& Sch\"onrich, R. 2016, \mnras, to appear (arXiv:1604.00191)

\bibitem[\protect\citeauthoryear{Aumer \& Sch\"onrich}{2015}]{AS15}
Aumer, M. \& Sch\"onrich, R. 2015, \mnras, {\bf 454}, 3166

\bibitem[\protect\citeauthoryear{Aumer, White \& Naab}{2014}]{AWN14}
Aumer, M., White, S. D. M. \& Naab, T. 2014, \mnras, {\bf 441}, 3679

\bibitem[\protect\citeauthoryear{Barazza \etal}{2009}]{Bara09}
Barazza, F. D., Jablonka, P., Desai, V.,\Omit{ Jogee, S., Arag\'On-Salamanca, A., De Lucia, G., Saglia, R. P., Halliday, C., Poggianti, B. M., Dalcanton, J. J., Rudnick, G., Milvang-Jensen, B., Noll, S., Simard, L., Clowe, D. I., Pell\'o, R., White, S. D. M. \& Zaritsky, D.} 2009, \aap, {\bf 497}, 713

\bibitem[\protect\citeauthoryear{Berrier \& Sellwood}{2015}]{BS15}
Berrier, J. \& Sellwood, J. A. 2015, \apj, {\bf 799}, 213

\bibitem[\protect\citeauthoryear{Bett \etal}{2007}]{Bett07}
Bett, P., Eke, V., Frenk, C. S., Jenkins, A., Helly, J. \& Navarro, J. 2007, \mnras, {\bf 376}, 215

\bibitem[\protect\citeauthoryear{Binney \& Tremaine}{2008}]{BT08}
Binney, J. \& Tremaine, S. 2008, {\it Galactic Dynamics\/} (2nd ed.; Princeton: Princeton University Press)

\bibitem[\protect\citeauthoryear{Bird \etal}{2013}]{Bird13}
Bird, J. C., Kazantzidis, S., Weinberg, D. H., Guedes, J., Callegari, S., Mayer, L. \& Madau, P. 2013, \apj, {\bf 773}, 43

\bibitem[\protect\citeauthoryear{Block \& Wainscoat}{1991}]{BW91}
Block, D. L. \& Wainscoat, R. J. 1991, \nat, {\bf 353}, 48

\bibitem[\protect\citeauthoryear{Blumenthal \etal}{1986}]{BFFP86}
Blumenthal, G. R., Faber, S. M., Flores, R. \& Primack, J. R. 1986, \apj, {\bf 301}, 27

\bibitem[\protect\citeauthoryear{Bosma}{1996}]{Bosm96}
Bosma, A. 1996, in {\it Barred Galaxies \& Circumnuclear Activity}, eds.\ Sandqvist, Aa.\ \& Lindblad, P. O. (Heidelberg: Springer) p~67

\bibitem[\protect\citeauthoryear{Bournaud \etal}{2005}]{BCS05}
Bournaud, F., Combes, F. \& Semelin, B. 2005, \mnras, {\bf 364}, L18

\bibitem[\protect\citeauthoryear{Bullock \etal}{2001}]{Bull01}
Bullock, J. S., Kolatt, T. S., Sigad, Y.,\Omit{ Somerville, R. S., Kravtsov, A. V., Klypin, A. A., Primack, J. R. \& Dekel, A.} 2001, \mnras, {\bf 321}, 559

\bibitem[\protect\citeauthoryear{Buta \& Block}{2001}]{BB01}
Buta, R. \& Block, D. L. 2001, \apj, {\bf 550}, 243

\bibitem[\protect\citeauthoryear{Buta \etal}{2015}]{Buta15}
Buta, R. J., Sheth, K., Athanassoula, E.,\Omit{ Bosma, A., Knapen, J. H., Laurikainen, E., Salo, H., Elmegreen, D., Ho, L. C., Zaritsky, D., Courtois, H., Hinz, J. L., Mu\~noz-Mateos, J.-C., Kim, T., Regan, M. W., Gadotti, D. A., Gil de Paz, A., Laine, J., Men\'endez-Delmestre, K., Comer\'on, S., Erroz Ferrer, S., Seibert, M., Mizusawa, T., Holwerda, B. \& Madore, B. F.} 2015, \apjs, {\bf 217}, 32

\bibitem[\protect\citeauthoryear{Byrd \etal}{1986}]{Byrd86}
Byrd, G. G., Valtonen, M. J., Valtaoja, L. \& Sundelius, B. 1986, \aap, {\bf 166}, 75

\bibitem[\protect\citeauthoryear{Cheung \etal}{2013}]{Cheu13}
Cheung, E., Athanassoula, E., Masters, K. L.,\Omit{ Nichol, R. C., Bosma, A., Bell, Eric F., Faber, S. M., Koo, D. C., Lintott, C., Melvin, T., Schawinski, K., Skibba, R. A. \& Willett, K. W.} 2013, \apj, {\bf 779}, 162

\bibitem[\protect\citeauthoryear{Christodoulou \etal}{1995}]{CST95}
Christodoulou, D. M., Shlosman, I. \& Tohline, J. E. 1995, \apj, {\bf 443}, 551

\bibitem[\protect\citeauthoryear{Col\'\i n \etal}{2006}]{CVK06}
Col\'\i n, P., Valenzuela, O. \& Klypin, A. 2006, \apj, {\bf 644}, 687

\bibitem[\protect\citeauthoryear{Courteau \etal}{2003}]{Cour03}
Courteau, S., Andersen, D. R., Bershady, M. A., MacArthur, L. A. \& Rix, H-W. 2003, \apj, {\bf 594}, 208

\bibitem[\protect\citeauthoryear{Danovich \etal}{2015}]{Dano15}
Danovich, M., Dekel, A., Hahn, O., Ceverino, D. \& Primack, J. 2015, \mnras, {\bf 449}, 2087

\bibitem[\protect\citeauthoryear{Davis \etal}{2012}]{Davi12}
Davis, B. L., Berrier, J. C., Shields, D. W.,\Omit{ Kennefick, J., Kennefick, D., Seigar, M. S., Lacy, C. H. S. \& Puerari, I.} 2012, \apjs, {\bf 199}, 33

\bibitem[\protect\citeauthoryear{Debattista \& Sellwood}{2000}]{DS00}
Debattista, V. P. \& Sellwood, J. A. 2000, \apj, {\bf 543}, 704

\bibitem[\protect\citeauthoryear{Efstathiou}{1979}]{Efst79}
Efstathiou, G. 1979, \mnras, {\bf 187}, 117

\bibitem[\protect\citeauthoryear{Efstathiou \etal}{1982}]{ELN82}
Efstathiou, G., Lake, G. \& Negroponte, J. 1982, \mnras, {\bf 199}, 1069

\bibitem[\protect\citeauthoryear{Elmegreen \etal}{1990}]{Elme90}
Elmegreen, D. M., Elmegreen, B. G. \& Bellin, A. D. 1990, \apj, {\bf 364}, 415

\bibitem[\protect\citeauthoryear{Erwin}{2005}]{Erwi05}
Erwin, P. 2005, \mnras, {\bf 364}, 283

\bibitem[\protect\citeauthoryear{Eskridge \etal}{2000}]{Eskr00}
Eskridge, P. B., Frogel, J. A., Pogge, R. W.,\Omit{ Quillen, A. C., Davies, R. L., DePoy, D. L., Houdashelt, M. L., Kuchinski, L. E., Ram\'\i rez, S. V., Sellgren, K., Terndrup, D. M. \& Tiede, G. P.} 2000, \aj, {\bf 119}, 536.

\bibitem[\protect\citeauthoryear{Gajda \etal}{2016}]{Gajd16}
Gajda, G., \L okas, E. L. \& Athanassoula, E. 2016, 2016, arXiv:1606.00322

\bibitem[\protect\citeauthoryear{Gerin \etal}{1990}]{Geri90}
Gerin, M., Combes, F. \& Athanassoula, E. 1990, \aap, {\bf 230}, 37

\bibitem[\protect\citeauthoryear{Grogin \etal}{2011}]{Grog11}
Grogin, N. A., Kocevski, D. D., Faber, S. M.,\Omit{ Ferguson, H. C., Koekemoer, A. M., Riess, A. G., Acquaviva, V., Alexander, D. M., Almaini, O., Ashby, M. L. N., Barden, M., Bell, E. F., Bournaud, F., Brown, T. M., Caputi, K. I., Casertano, S., Cassata, P., Castellano, M., Challis, P., Chary, R.-R., Cheung, E/, Cirasuolo, M., Conselice, C. J., Roshan Cooray, A., Croton, D. J., Daddi, E., Dahlen, T., Dav\'e, R., de Mello, D. F., Dekel, A., Dickinson, M., Dolch, T., Donley, J. L., Dunlop, J. S., Dutton, A. A., Elbaz, D., Fazio, G. G., Filippenko, A. V., Finkelstein, S. L., Fontana, A., Gardner, J. P., Garnavich, P. M., Gawiser, E., Giavalisco, M., Grazian, A., Guo, Y., Hathi, N. P., H\"aussler, B., Hopkins, P. F., Huang, J-S., Huang, K-H., Jha, S. W., Kartaltepe, J. S., Kirshner, R. P., Koo, D. C., Lai, K., Lee, K-S., Li, W., Lotz, J. M., Lucas, R. A., Madau, P., McCarthy, P. J., McGrath, E. J., McIntosh, D. H., McLure, R. J., Mobasher, B., Moustakas, L. A., Mozena, M., Nandra, K., Newman, J. A., Niemi, S-M., Noeske, K. G., Papovich, C. J., Pentericci, L., Pope, A., Primack, J. R., Rajan, A., Ravindranath, S., Reddy, N. A., Renzini, A., Rix, H-W., Robaina, A. R., Rodney, S. A., Rosario, D. J., Rosati, P., Salimbeni, S., Scarlata, C., Siana, B., Simard, L., Smidt, J., Somerville, R. S., Spinrad, H., Straughn, A. N., Strolger, L-G., Telford, O., Teplitz, H. I., Trump, J. R., van der Wel, A., Villforth, C., Wechsler, R. H., Weiner, B. J., Wiklind, T., Wild, V., Wilson, G., Wuyts, S., Yan, H-J. \& Yun, M. S.} 2011, \apjs, {\bf 197}, 35

\bibitem[\protect\citeauthoryear{Guedes \etal}{2011}]{Gued11}
Guedes, J., Callegari, S., Madau, P. \& Mayer, L. 2011, \apj, {\bf 742}, 76

\bibitem[\protect\citeauthoryear{Hart \etal}{2016}]{Hart16}
Hart, R. E., Bamford, S. P., Willett, K. W.,\Omit{ Masters, K. L., Cardamone, C., Lintott, C. J., Mackay, R. J., Nichol, R. C., Rosslowe, C. K., Simmons, B. D. \& Smethurst, R. J.} 2016, arXiv:1607.01019

\bibitem[\protect\citeauthoryear{Hernquist}{1990}]{Hern90}
Hernquist, L. 1990, \apj, {\bf 356}, 359

\bibitem[\protect\citeauthoryear{Koekemoer \etal}{2011}]{Koek11}
Koekemoer, A. M., Faber, S. M., Ferguson, H. C.,\Omit{ Grogin, N. A., Kocevski, D. D., Koo, D. C., Lai, K., Lotz, J. M., Lucas, R. A., McGrath, E. J., Ogaz, S., Rajan, A., Riess, A. G., Rodney, S. A., Strolger, L., Casertano, S., Castellano, M., Dahlen, T., Dickinson, M., Dolch, T., Fontana, A., Giavalisco, M., Grazian, A., Guo, Y., Hathi, N. P., Huang, K-H., van der Wel, A., Yan, H-J., Acquaviva, V., Alexander, D. M., Almaini, O., Ashby, M. L. N., Barden, M., Bell, E. F., Bournaud, F., Brown, T. M., Caputi, K. I., Cassata, P., Challis, P. J., Chary, R-R., Cheung, E., Cirasuolo, M., Conselice, C. J., Roshan Cooray, A., Croton, D. J., Daddi, E., Dav\'e, R., de Mello, D. F., de Ravel, L., Dekel, A., Donley, J. L., Dunlop, J. S., Dutton, A. A., Elbaz, D., Fazio, G. G., Filippenko, A. V., Finkelstein, S. L., Frazer, C., Gardner, J. P., Garnavich, P. M., Gawiser, E., Gruetzbauch, R., Hartley, W. G., H\"aussler, B., Herrington, J., Hopkins, P. F., Huang, J-S., Jha, S. W., Johnson, A., Kartaltepe, J. S., Khostovan, A. A., Kirshner, R. P., Lani, C., Lee, K-Soo., Li, W., Madau, P., McCarthy, P. J., McIntosh, D. H., McLure, R. J., McPartland, C., Mobasher, B., Moreira, H., Mortlock, A., Moustakas, L. A., Mozena, M., Nandra, K., Newman, J. A., Nielsen, J. L., Niemi, S., Noeske, K. G., Papovich, C. J., Pentericci, L., Pope, A., Primack, J. R., Ravindranath, S., Reddy, N. A., Renzini, A., Rix, H-W., Robaina, A. R., Rosario, D. J., Rosati, P., Salimbeni, S., Scarlata, C., Siana, B., Simard, L., Smidt, J., Snyder, D., Somerville, R. S., Spinrad, H., Straughn, A. N., Telford, O., Teplitz, H. I., Trump, J. R., Vargas, C., Villforth, C., Wagner, C. R., Wandro, P., Wechsler, R. H., Weiner, B. J., Wiklind, T., Wild, V., Wilson, G., Wuyts, S. \& Yun, Min S.} 2011, \apjs, {\bf 197}, 36

\bibitem[\protect\citeauthoryear{Kormendy}{2013}]{Korm13}
Kormendy, J. 2013, In XXIII Canary Islands Winter School of Astrophysics, ``Secular Evolution of Galaxies'' eds. J. Falc\'on-Barroso \& J. H. Knapen (Cambridge: Cambridge University Press), p.~1

\bibitem[\protect\citeauthoryear{Kormendy \etal}{2010}]{Korm10}
Kormendy, J., Drory, N., Bender, R. \& Cornell, M. E. 2010, \apj, {\bf 723}, 54

\bibitem[\protect\citeauthoryear{Kormendy \& Kennicutt}{2004}]{KK04}
Kormendy, J. \& Kennicutt, R. C. 2004, \araa, {\bf 42}, 603

\bibitem[\protect\citeauthoryear{Li \etal}{2009}]{Li09}
Li, C., Gadotti, D. A., Mao, S. \& Kauffmann, G. 2009, \mnras, {\bf 397}, 726

\bibitem[\protect\citeauthoryear{Lin \etal}{2014}]{Lin14}
Lin, Y., Cervantes Sodi, B., Li, C., Wang, L. \& Wang, E. 2014, \apj, {\bf 796}, 98

\bibitem[\protect\citeauthoryear{Lowing \etal}{2015}]{Lowi15}
Lowing, B., Wang, W., Cooper, A., Kennedy, R., Helly, J., Cole, S. \& Frenk, C. 2015, \mnras, {\bf 446}, 2274

\bibitem[\protect\citeauthoryear{Lovelace \& Hohlfeld}{1978}]{LH78}
Lovelace, R. V. E. \& Hohlfeld, R. G. 1978, \apj, {\bf 221}, 51

\bibitem[\protect\citeauthoryear{Marinova \& Jogee}{2007}]{MJ07}
Marinova, I. \& Jogee, S. 2007, \apj, {\bf 659}, 1176

\bibitem[\protect\citeauthoryear{Marinova \etal}{2012}]{Mari12}
Marinova, I., Jogee, S., Weinzirl, T.,\Omit{ Erwin, P., Trentham, N., Ferguson, H. C., Hammer, D., den Brok, M., Graham, A. W., Carter, D., Balcells, M., Goudfrooij, P., Guzm\'an, R., Hoyos, C., Mobasher, B., Mouhcine, M., Peletier, R. F., Peng, E. W. \& Verdoes Kleijn, G.} 2012, \apj, {\bf 746}, 136

\bibitem[\protect\citeauthoryear{Martinez-Valpuesta, Shlosman \& Heller}{2006}]{MVSH}
Martinez-Valpuesta, I., Shlosman, I. \& Heller, C. 2006, \apj, {\bf 637}, 214

\bibitem[\protect\citeauthoryear{Masters \etal}{2011}]{Mast11}
Masters, K. L., Nichol, R. C., Hoyle, B.,\Omit{ Lintott, C., Bamford, S., Edmondson, E. M., Fortson, L., Keel, W. C., Schawinski, K., Smith, A. \& Thomas, D.} 2011, \mnras, {\bf 411}, 2026

\bibitem[\protect\citeauthoryear{Masters \etal}{2012}]{Mast12}
Masters, K. L., Nichol, R. C., Haynes, M. P.,\Omit{ Keel, W. C., Lintott, C., Simmons, B., Skibba, R., Bamford, S., Giovanelli, R. \& Schawinski, K.} 2012, \mnras, {\bf 424}, 2180

\bibitem[\protect\citeauthoryear{Mathewson \& Ford}{1996}]{MF96}
Mathewson, D. S. \& Ford, V. L. 1996, \apjs, {\bf 109}, 97

\bibitem[\protect\citeauthoryear{Matteucci \& Francois}{1989}]{MF89}
Matteucci, F. \& Francois, P. 1989, \mnras, {\bf 239}, 885

\bibitem[\protect\citeauthoryear{Mayer \& Wadsley}{2004}]{MW04}
Mayer, L. \& Wadsley, J. 2004, \mnras, {\bf 347}, 277

\bibitem[\protect\citeauthoryear{Munshi \etal}{2013}]{Muns13}
Munshi, F., Governato, F., Brooks, A. M.,\Omit{ Christensen, C., Shen, S., Loebman, S., Moster, B., Quinn, T. \& Wadsley, J.} 2013, \apj, {\bf 766}, 56

\bibitem[\protect\citeauthoryear{Noguchi}{1987}]{Nogu87}
Noguchi, M. 1987, \mnras, {\bf 228}, 635

\bibitem[\protect\citeauthoryear{Norman \etal}{1996}]{NSH96}
Norman, C. A., Sellwood, J. A. \& Hasan, H. 1996, \apj, {\bf 462}, 114

\bibitem[\protect\citeauthoryear{Pfenniger \& Norman}{1990}]{PF90}
Pfenniger, D. \& Norman, C. 1990, \apj, {\bf 363}, 391

\bibitem[\protect\citeauthoryear{Ostriker \& Peebles}{1973}]{OP73}
Ostriker, J. P. \& Peebles, P. J. E. 1973, \apj, {\bf 186}, 467

\bibitem[\protect\citeauthoryear{Polyachenko \etal}{2016}]{PBJ16}
Polyachenko, E., Berczik, P. \& Just, A. 2016, \mnras, subiteed (arXiv:1601.06115)

\bibitem[\protect\citeauthoryear{Reese \etal}{2007}]{Rees07}
Reese, A., Williams, T. B., Sellwood, J. A., Barnes, E. I. \& Powell, B. A. 2007, \aj, {\bf 133}, 2846

\bibitem[\protect\citeauthoryear{Romano-D\'\i az \etal}{2008}]{Roma08}
Romano-D\'\i az, E., Shlosman, I., Heller, C. \& Hoffman, Y. 2008, \apjl, {\bf 687}, L13

\bibitem[\protect\citeauthoryear{Saha \& Naab}{2013}]{SN13}
Saha, K. \& Naab, T. 2013, \mnras, {\bf 434}, 1287

\bibitem[\protect\citeauthoryear{Salo}{1991}]{Salo91}
Salo, H. 1991, \aap, {\bf 243}, 118

\bibitem[\protect\citeauthoryear{Schaller \etal}{2015}]{Scha15}
Schaller, M., Frenk, C. S., Bower, R. G.,\Omit{ Theuns, T., Jenkins, A., Schaye, J., Crain, R. A., Furlong, M., Dalla Vecchia, C. \& McCarthy, I. G.} 2015, \mnras, {\bf 451}, 1247

\bibitem[\protect\citeauthoryear{Sellwood}{1989}]{Sell89}
Sellwood, J. A. 1989, \mnras, {\bf 238}, 115

\bibitem[\protect\citeauthoryear{Sellwood}{2003}]{Sell03}
Sellwood, J. A. 2003, \apj, {\bf 587}, 638

\bibitem[\protect\citeauthoryear{Sellwood}{2013}]{Sell13}
Sellwood, J. A. 2013, in {\it Planets Stars and Stellar Systems}, v.{\bf 5}, eds.\ T. Oswalt \& G. Gilmore (Heidelberg: Springer), 923 (arXiv:1006.4855)

\bibitem[\protect\citeauthoryear{Sellwood}{2014}]{Sell14}
Sellwood, J. A. 2014, arXiv:1406.6606 (on-line manual: \hfil\break {\tt http://www.physics.rutgers.edu/$\sim$sellwood/manual.pdf})

\bibitem[\protect\citeauthoryear{Sellwood}{2016}]{Sell16}
Sellwood, J. A. 2016, \apj, {\bf 819}, 92

\bibitem[\protect\citeauthoryear{Sellwood \& Binney}{2002}]{SB02}
Sellwood, J. A. \& Binney, J. J. 2002, \mnras, {\bf 336}, 785

\bibitem[\protect\citeauthoryear{Sellwood \& Carlberg}{1984}]{SC84}
Sellwood, J. A. \& Carlberg, R. G. 1984, \apj, {\bf 282}, 61

\bibitem[\protect\citeauthoryear{Sellwood \& Debattista}{2006}]{SD06}
Sellwood, J. A. \& Debattista, V. P. 2006, \apj, {\bf 639}, 868

\bibitem[\protect\citeauthoryear{Sellwood \& Evans}{2001}]{SE01}
Sellwood, J. A. \& Evans, N. W. 2001, \apj, {\bf 546}, 176

\bibitem[\protect\citeauthoryear{Sellwood \& McGaugh}{2005}]{SM05}
Sellwood, J. A. \& McGaugh, S. S. 2005, \apj, {\bf 634}, 70

\bibitem[\protect\citeauthoryear{Sellwood \& Moore}{1999}]{SM99}
Sellwood, J. A. \& Moore, E. M. 1999, \apj, {\bf 510}, 125

\bibitem[\protect\citeauthoryear{Shen \& Sellwood}{2004}]{SS04}
Shen, J. \& Sellwood, J. A. 2006, \mnras, {\bf 370}, 2

\bibitem[\protect\citeauthoryear{Silk \& Mamon}{2012}]{SM12}
Silk, J. \& Mamon, G. A. 2012, \raa, {\bf 12}, 917

\bibitem[\protect\citeauthoryear{Skibba \etal}{2012}]{Skib12}
Skibba, R. A., Masters, K. L., Nichol, R. C.,\Omit{ Zehavi, I., Hoyle, B., Edmondson, E. M., Bamford, S. P., Cardamone, C. N., Keel, W. C., Lintott, C. \& Schawinski, K.} 2012, \mnras, {\bf 423}, 1485

\bibitem[\protect\citeauthoryear{Snyder \etal}{2015}]{Snyd15}
Snyder, G. F., Torrey, P., Lotz, J. M.,\Omit{ Genel, S., McBride, C. K., Vogelsberger, M., Pillepich, A., Nelson, D., Sales, L. V., Sijacki, D., Hernquist, L. \& Springel, V.} 2015, \mnras, {\bf 454}, 1886

\bibitem[\protect\citeauthoryear{Somerville \& Dav\'e}{2015}]{SD15}
Somerville, R. S. \& Dav\'e, R.	2015, \araa, {\bf 53}, 51

\bibitem[\protect\citeauthoryear{Somerville, Popping \& Trager}{2015}]{SPT15}
Somerville, R. S., Popping, G. \& Trager, S. C. 2015, \mnras, {\bf 453}, 4337

\bibitem[\protect\citeauthoryear{Stinson \etal}{2013}]{Stin13}
Stinson, G. S., Brook, C., Macci\`o, A. V., Wadsley, J., Quinn, T. R. \& Couchman, H. M. P. 2013, \mnras, {\bf 428}, 129

\bibitem[\protect\citeauthoryear{Toomre}{1963}]{Toom63}
Toomre, A. 1963, \apj, {\bf 138}, 385

\bibitem[\protect\citeauthoryear{Toomre}{1969}]{Toom69}
Toomre, A. 1969, \apj, {\bf 158}, 899

\bibitem[\protect\citeauthoryear{Toomre}{1981}]{Toom81}
Toomre, A. 1981, in {\it The Structure and Evolution of Normal Galaxies}, ed.\ S. M. Fall \& D. Lynden-Bell (Cambridge: Cambridge University Press), p.~111

\bibitem[\protect\citeauthoryear{Wang \etal}{2011}]{Wang11}
Wang, J., Kauffmann, G., Overzier, R.,\Omit{ Catinella, B., Schiminovich, D., Heckman, T. M., Moran, S. M., Haynes, M. P., Giovanelli, R. \& Kong, X.} 2011, \mnras, {\bf 412}, 1081

\bibitem[\protect\citeauthoryear{Weinberg \etal}{2015}]{Wein15}
Weinberg, D. H., Bullock, J. S., Governato, F., Kuzio de Naray, R. \& Peter, A. H. G. 2015, \pnas, {\bf 112}, 12249

\bibitem[\protect\citeauthoryear{Young}{1980}]{Youn80}
Young, P. 1980, \apj, {\bf 242}, 1232

\bibitem[\protect\citeauthoryear{Zang}{1976}]{Zang76}
Zang, T. A. 1976, \PhD, MIT


\end{thebibliography}
\end{document}